# Generation and Influence of Eccentric Ideas on Social Networks

Sriniwas Pandey[1, *], Yiding Cao[1], Yingjun Dong[2], Minjun Kim[1], Neil G. MacLaren[3], Shelley D. Dionne[1], Francis J. Yammarino[1], and Hiroki Sayama[1, 4, *]

[1]Binghamton University, NY, USA, [2]University of Texas Health Science Center at Houston, Texas, USA, [3]State University of New York at Buffalo, Buffalo, NY, USA, [4]Faculty of Commerce, Waseda University, Japan

## Abstract

Studying extreme ideas in routine choices and discussions is of utmost importance to understand the increasing polarization in society. In this study, we focus on understanding the generation and influence of extreme ideas in routine conversations which we label "eccentric" ideas. The eccentricity of any idea is defined as the deviation of that idea from the norm of the social neighborhood. We collected and analyzed data from two completely different sources: public social media and online experiments in a controlled environment. We compared the popularity of ideas against their eccentricity to understand individuals' fascination towards eccentricity. We found that more eccentric ideas have a higher probability of getting a greater number of "likes". Additionally, we demonstrate that the social neighborhood of an individual conceals eccentricity changes in one's own opinions and facilitates generation of eccentric ideas at a collective level.

## Main

With rapidly diminishing global boundaries, readily available communication devices, and increasing popularity of social media, a new trend of being "extreme" is becoming normal in today's attention-driven society. In domains like religion and politics, this trend is evident, whereas in areas like music, lifestyle, food, healthcare, and other day-to-day choices, it is not as visible but still present covertly. The ubiquitous presence of such trends makes it important to understand their causes and effects and identify methods to intervene when necessary and appropriate.

Religious and violent extremism has been a well-discussed topic in the literature due to their global presence and destructive outcomes [1–5]. However, there are other categories of extremism that grow gradually over time and remain unnoticed until their ultimate consequences are visible. We use the term "eccentricity" to distinguish such extremism from traditional political and religious extremism. The impact of eccentricity can be perceived in various forms. By spreading hate and disharmony, eccentric behavior polarizes society[6,7]. Fanaticism for favorite celebrities, singers, and politicians often leads to personal rivalry, threats, and cyber-bullying [8,9]. In some other scenarios, eccentricity can be more injurious and fatal too. Such outcomes have been observed in anti-vaccine movements[10–12] and firearms-related discourses[13,14]. The impact of eccentric opinions and behaviors is not only limited to social and individual issues, but it influences financial and economic domains too. Industries often utilize this behavioral tendency to change the dynamics of the market. The popularity of only vegetarian/non-vegetarian food[15,16], vegan diets, and GMO/anti-GMO foods[17] is the result of such exploitation by the related corporations and social organizations. These examples suggest that a systematic investigation and research on this category of extremism (eccentricity) is required.

Understanding the genesis and evolution of extremism has always been of great interest to researchers[18–22]. Factors like gender, race, education, and upbringing do affect the thought process of an individual[23–25]. However, the influence of society, social interactions, and information exposure plays a dominant role in building or altering opinions[21,26–32]. To understand the factors and underlying opinion dynamics researchers have developed different strategies[33,34]. Researchers have applied statistical physics and other mathematical models to examine the dynamics of extreme idea generation[35–37]. Simulation models like agent-based modeling have also been utilized in several studies[21,36,38,39]. In this study, we uniquely investigated this problem using multiple empirical data of human idea generation and evolution: experimental data obtained through human-subject experiments of idea generation[40,41] and online ideation and interaction data obtained from public social media GAB[42]. We analyzed both datasets and compared the results to identify common patterns of how eccentricity may arise in social networks. To our knowledge, this is the first study that investigated the eccentric idea generation and evolution dynamics using extensive real-world empirical data obtained from different sources.

To understand the inception and development of eccentricity, the first step is to establish a quantitative, measurable definition of eccentricity. There are several works that define traditional extremism, but these definitions are restricted to the application and domain of their work and cannot be directly used to define eccentricity in a broader context[43–45]. Each definition covers some aspects of extremism and ignores the rest[46,47]. As most researchers consider extremism in terms of violence, they weigh an opinion based on whether the opinion will result in an action or not[4,45]. If the opinion results in an action (mostly destructive), the opinion is considered extreme, e.g., those leading to terrorism or a coup. The major drawback of this popular definition is that it does not quantify the strength of opinions. As this definition makes a binary classification about extremism or not, it would miss the scenarios where extremism might not result in direct actions but may cause more gradual escalation of opinions.

Another common way to define eccentricity is using the threshold method in continuous models[18,19]. In this method, an opinion's strength is quantified in a range of possible values and if the strength is greater than a predefined threshold, that opinion is considered extreme or eccentric[26,48]. One of the major challenges with this method is deciding the threshold. The most appropriate threshold can vary for different environments, different backgrounds and different tasks, and the same opinion will not qualify as extreme in one setting versus another. Even in the same setting, with time, the opinions that were initially classified eccentric may sound sensible in presence of more eccentric ideas later, and vice versa.

In our study, we chose the simplest dictionary definition of eccentricity: "deviation from the norm". The norm of a conversation or discussion is defined as the center of all opinions in a social neighborhood, and the eccentricity is quantified as the distance from the norm, both in a semantic metric space that is obtained using machine learning techniques for semantic embedding. This approach makes our definition of eccentricity parameter-free and avoids the problem of binary classification. As the metric is continuous and context-free, the method can be applied to any domain or task without the need for domain-specific knowledge or expertise.

***Ideas that become more popular are intrinsically more eccentric.*** We focus on understanding the influence of the eccentricity of an idea on the amount of attention the idea receives. From two different data sources (human-subject experiments[40,41] and online social media GAB[42]), we collected three types of data: 1) text posts collected from an online experiment for a laptop tagline writing (high collaboration) task, 2) text posts collected from an online experiment for story writing (low collaboration) task, and 3) social media posts from GAB (see Methods for details). The number of likes on a post recorded in these datasets is considered a representation of the amount of attention the post received. The eccentricity of a posted opinion is measured by the semantic distance between the idea and the center of all other ideas within the individual's social neighborhood (Figure 1).

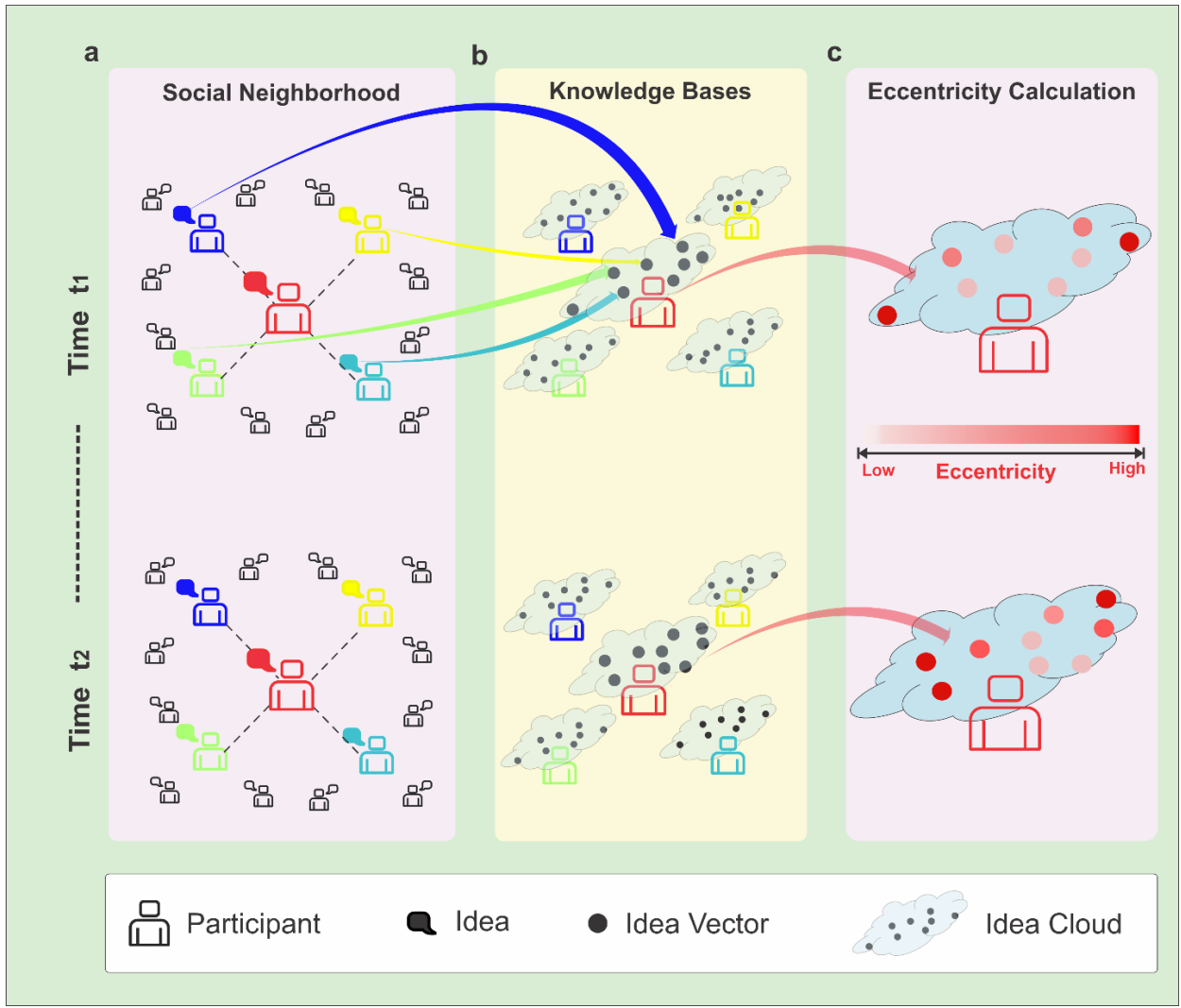

**Figure 1: Schematic illustration of idea eccentricity calculation**. Each participant is connected to several other participants (neighbors) in a social network. **(a)** Each participant can see the ideas of his or her neighbors. **(b)** Ideas recently posted by the participant and her neighbors constitute her *Knowledge Base* at each point in time and those ideas are vectorized using Doc2Vec to form an *Idea Cloud*. Every participant has a separate knowledge base, which is updated whenever a new idea is posted in the neighborhood. **(c)** The distance of an idea vector from the center of the idea cloud is the measure of the *eccentricity* of that idea.

The social neighborhood of an individual, $U_i$, is his or her ego network, which comprises all other individuals $U_i$ is following and user $U_i$ herself. $U_i$ gets exposed to the opinions posted in her social neighborhood. Recent opinions posted (we used recent 5 days) in the social neighborhood of $U_i$ constitute a knowledge base of $U_i$ at each time point $t$. The notion behind maintaining a knowledge base is that the exposure to recent ideas might have an impact on new ideas. In a long period of time, the topics and discussions can change drastically and will not have much impact on new opinions. Text ideas are converted into numerical vectors using the Doc2Vec method[49]. Principal component analysis is performed to reduce the dimensionality of these numerical vectors (see Methods for details). Ideas in the knowledge base, when represented in numerical vector form, create an idea cloud, $C_i$, for user $U_i$. The knowledge base and the idea cloud are different for each individual and get updated whenever there is a new idea added in the neighborhood. The eccentricity of a new idea $P_i$, posted by $U_i$, is measured by the distance of $P_i$ from the center of the idea cloud $C_i$ at that time point.

Posted ideas are partitioned into different popularity levels according to the number of likes they received. We compare the probability distributions of eccentricity for different popularity levels to find the relationship between eccentricity and popularity. We use the kernel density estimation method with a Gaussian filter[50] to estimate the probability distribution for each popularity level. Figures 2a and 2b represent popularity distributions for the posts in the laptop tagline writing task and the posts in the short story writing task, respectively.

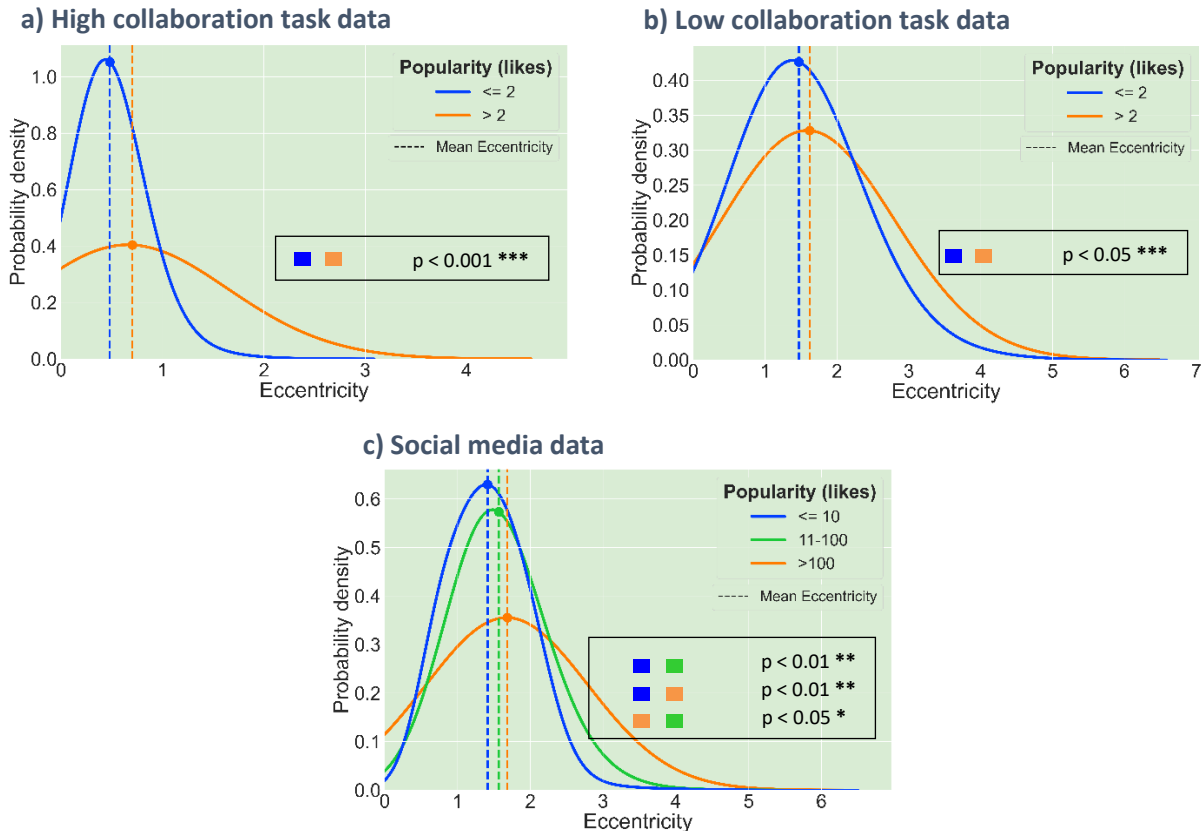

**Figure 2. Probability distributions of the eccentricity of the posted ideas at different popularity levels (number of likes, i.e., amount of attention they attracted).** Popularity levels are represented in different color bins, low (blue) and high (orange) (and, in case of GAB data, medium (green)). The vertical dashed lines show the average value of eccentricity for each popularity level. **a.** Plots for the dataset collected in the laptop tagline writing experiment (high collaboration task). Two-sample Anderson-Darling test with Bonferroni correction on unequal sample sizes ($n_{>2}$ = 59, $n_{<=2}$ = 815) b. Plots for the dataset collected in the

short story writing experiment (low collaboration task). Two-sample Anderson-Darling test with Bonferroni correction on unequal sample sizes ($n_{<=2}$ = 669, $n_{>2}$ = 38) **c.** Plots for the dataset collected from GAB. Two-sample Anderson-Darling test with Bonferroni correction on unequal sample sizes ($n_{<=10}$ = 130234, $n_{11-100}$ = 1866, $n_{>100}$90). In the plots of high popularity levels, the tail of the probability density function becomes broader in all three data sources. The average eccentricity also increases as the popularity level goes up in all cases.

The data obtained from the online experiments has a limited range of likes (0 - 5), hence ideas are partitioned into two popularity levels: **High** (>2) and **Low** (<=2). Figure 2c shows the probability distribution for GAB social media posts. GAB posts, which have a wide range of the number of likes (0 to 500+), are partitioned into three popularity levels: **Low** (<=10), **Medium** (11-100), and **High** (>100). For all the three datasets, the right tail of the distribution gets broader for the higher popularity levels, indicating that more eccentric posts attract greater attention and popularity. The pattern of increasing average eccentricity with increasing popularity levels is consistent across all the datasets, despite the different nature of the sources. The average eccentricity and eccentricity distributions for different popularity levels are significantly different from each other (p-values obtained by Anderson-Darling test using Bonferroni correction method are shown on each plot).

***Movement of neighborhood ideas conceals own ideas' deviation:*** Furthermore, we propose another type of eccentricity measure named "self-eccentricity". Self-eccentricity is the eccentricity of opinions with respect to the previous opinions made by the same author of the opinion in question. In other words, the self-eccentricity of an individual's new idea is the distance of the new idea from the center of that individual's previous ideas. Whereas eccentricity is an indicator of deviation from the common consensus or core of the discussion in the social neighborhood at a certain time point, self-eccentricity measures departure from one's own previous ideas over time. We applied this measure to the GAB dataset that had sufficient historical data of users' opinions. We observed that the eccentricities of posts made by an individual did not remain constant but kept changing. We quantified the temporal change of self-eccentricity and analyzed the dynamics of eccentricity change for each user.

To quantify the overall change in eccentricity of an individual, we use the F-score and G-score metric proposed by Mall et.al[51]. F-score is a weighted average of absolute change of eccentricity (without considering the direction). G-score takes into consideration the direction of change of eccentricity as well. The time duration between two consecutive ideas and the average time duration between ideas is also included in the calculation. F-score and G-score combined characterize the behavior of an individual. Additionally, we calculate F-score and G-score based on the self-eccentricity for each user. F-score and G-score of self-eccentricity combined characterize the deviation from one's own ideas in the past.

Each individual is represented by F-score and G-score on a 2-D space in Figure 3a. The distribution is symmetrical in the positive and negative sides of the G-score and most of the individuals lie near the G-score zero line, implying that there is no particular directional trend in idea eccentricity with regard to the user's social neighborhood. In other words, an individual's ideas are generally more or less consistent with those in his or her social neighborhood's. However, when users are placed on their self-eccentricity's F-score and G-score space, the plot is clearly skewed towards positive G-scores (Figure 3b). This means that users are generally deviating away from their previous ideas and thus their self-eccentricity increases over time in terms of their previous opinions, regardless of their eccentricity in their social neighborhood (as shown by the colors of markers).

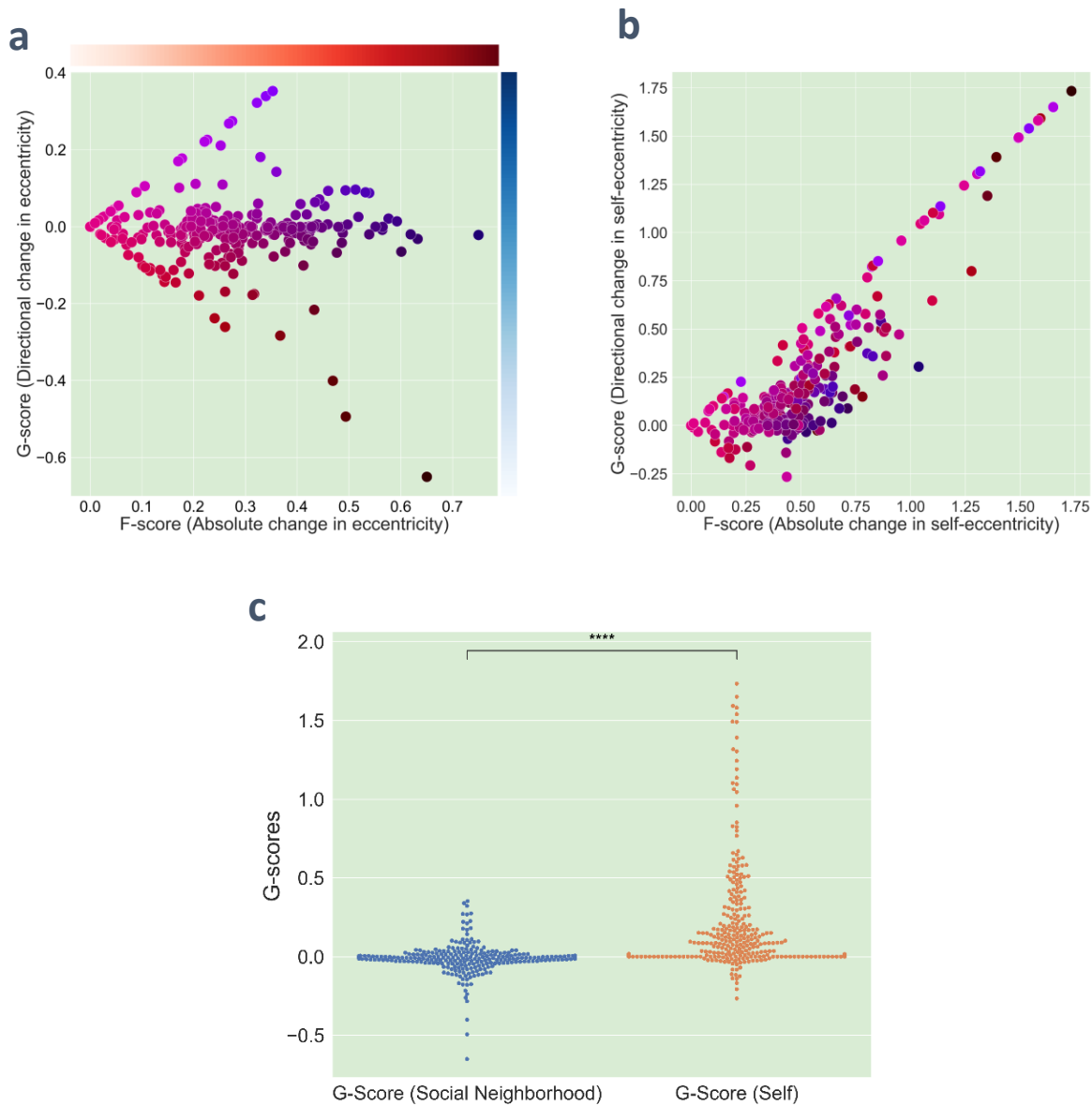

**Figure 3. Distribution of GAB users' idea eccentricity dynamics.** Each user's dynamics are quantified using the F-score (weighted average of absolute change; horizontal axis) and g-score (directional change; vertical axis) [44] of his or her idea eccentricities. Each marker represents one GAB user. **a.** F-scores and G-scores of users' eccentricities in their respective social neighborhood, i.e., deviation of their ideas from the center of their neighbors' ideas. Users are colored according to their (F-score, G-score) coordinates as shown in color bars. **b.** F-scores and G-scores of the same users' self-eccentricities, i.e., deviation of their ideas from their own past ideas. Each user in (b) is colored using the same color as in (a) to show the correspondence between the two plots. **c**. G-score distributions for eccentricity relative to the social neighborhood and self-eccentricity, which are significantly different (Mann-Whitney test with equal sample size of 318) from each other with G-score (social neighborhood) mean lower than the G-score (self) mean (- 0.0092 vs 0.201)

These two results, when interpreted together, deliver the following key finding of our work: Individuals are turning more eccentric with time in terms of their own previous ideas (Figure 3b); however, as everyone in the neighborhood of these individuals is also shifting from their prior opinions, the change in individual's ideas may not be noticeable (Figure 3a). Such spontaneous yet unrecognized increase of idea eccentricity can be driven by the positive correlation between eccentricity and attention described earlier (Figure 2).

## Discussion

In today's heavily interconnected world, we notice a trend that a large section of society is increasingly opting for eccentric choices that stand out. We have explored a few insights about this behavior with multiple real-world empirical data. Our first finding shows that the deviation of opinions from the norm helps attract the attention of other individuals. Several studies have shown that serious adverse effects of social media usage are cravings to receive social acceptance and attract the attention of friends and acquaintances[52–54]. Our finding indicates that such human desire may naturally lead to generation of more eccentric opinions. This behavior may scale up to other contexts in the real world beyond online social media.

Another crucial implication about eccentricity obtained in this study is that the overall collective shift of ideas in our social neighborhood may create an illusion of consistency in our own opinions. This can be understood using an analogy of multiple passengers riding on an elevator. In a smooth-moving elevator, any change in elevation is not felt directly by the people using the elevator because the only reference points to assess one's position are fellow individuals riding on the same elevator. As everyone is moving in the same direction at the same speed, it feels like everyone is standing still and not moving. Similarly, we may not feel the shift in eccentricity of our opinions as our social neighbors show similar shifts.

This study draws a picture of how extreme ideas and opinions may spontaneously arise in society. Everyone wants to gain social acceptance and become popular and influential. As being eccentric in opinions helps attract neighbors' attention, people start expressing out-of-center opinions. And as most of the social neighbors do the same, it would be difficult to notice that one's opinions are becoming more eccentric compared to others. If people do not recognize the heat, there would be little feedback mechanisms to stop them from becoming more eccentric. These behavioral patterns form a cycle and may reinforce each other. These conclusions illustrate the need for further study of how to detect such spontaneous escalation dynamics in society, and if appropriate, how to implement effective interventions so that it won't cause undesirable negative impacts on our lives.

## Methods

***Online experiments and data collection:*** We collected data from two different sources. One is a human-subject online experiments performed at a mid-sized US university, where students in different majors were recruited to participate in a collaborative textual design task on a Twitter-like online experimental platform[40,41]. Participants were linked to a subset of other participants like a social network setting. There were two types of tasks used for the experiment: (i) writing laptop marketing taglines (high collaboration task) and (ii) writing short fictional stories (low collaboration task). Like in typical social media platforms, participants in these experiments could see only the ideas posted by their social neighbors, like those neighbors' ideas, and add comments to them. Details of the experiments, idea generation process and method of visualizations can be found in these papers[40,41].

The other data source we utilized for data collection is the GAB social media. GAB is a social networking service particularly popular among the far-right people in the US. The data from the GAB is freely available and can be scraped from their website. We used the snowball sampling method to collect data from GAB. We collected around 30M posts and connections information between authors of these posts. As we are interested in understanding eccentricity, which requires a connected network of individuals, we selected the largest connected component (LCC) from the user network. A subgraph induced by randomly selected 10% of the users in the LCC was used for data analysis to keep the computational demand at a manageable level. This gave us a dataset of around 3,000 GAB users with about 147,000 posts for the analysis. The dataset consists of posts that were made between August 2016 and January 2021.

***Text embedding and dimensionality reduction:*** The first step in analyzing textual data is to convert it into numerical form, a process called text embedding. Before converting text ideas into numerical vectors, each text idea was cleaned to remove stop words, punctuation marks and digits. We also used word stemming to convert different forms of a word to a standard form. In the subsequent step, the Doc2Vec method was used to convert the cleaned text ideas into numerical vectors. Doc2Vec first creates a vocabulary using the text corpus (all ideas combined in this study), trains a model and infers a numerical vector for each text idea. For the GAB data, we set the inferred vector size to be 300, whereas for online experiment data it was set to 90, given the difference in the data sizes. The 300-dimensional numerical vectors for the GAB data were further transformed to lower dimensional vectors (115 dimensions) using the principal component analysis preserving 90% variance in the data. We call the resulting numerical vector an "idea vector."

***Eccentricity and self-eccentricity calculation:*** To calculate eccentricity of an idea, we first create a social neighborhood for each user, which is a directed network between a user and all other users he or she is following. In our study, we assume the collaboration network remains unchanged for each user during the study. The collaboration network is used to create a knowledge base for each user. A knowledge base is the collection of recent ideas (recent 5 days) made in the neighborhood. An idea cloud of a user is the collection of his or her idea vectors in a multi-dimensional space. The mean vector of an idea cloud is the center, and the distance ($L^2$ norm was used in this study) of an idea vector from the center of the idea cloud is the eccentricity of that idea. One important point to remember here is that the knowledge base and idea clouds are different for every individual and get updated whenever there is a new idea in the neighborhood. Eccentricity is defined according to the current idea cloud of a particular user.

Self-eccentricity is the deviation from the center of one's own recent previous ideas (recent 5 days). For each individual, a history of recent ideas is maintained, and each new idea is evaluated against the history. The distance of a new idea from the center of the history cloud for the author of that idea is a measure of self-eccentricity of the new idea.

***Popularity levels and eccentricity distribution:*** Ideas are categorized into different popularity levels based on the number of likes they received. The online experiments had a limited number of participants, and each session ran only for two weeks, so the maximum number of likes is less than 6 for this dataset (laptop taglines data < 5, story writing data < 6)[40,41]. As the range of number of likes is narrow, the ideas from online experiments are divided into just two categories: ideas having fewer than or equal to two likes (Low popularity) and ideas having more than two likes (High popularity). In case of the GAB dataset, the range of number of likes is much wider (0 to 500+), hence we partition ideas into three categories: fewer than or equal to ten likes (Low popularity), eleven to one hundred likes (Medium popularity) and more than one hundred likes (High popularity). For each popularity level, a probability distribution of eccentricity is constructed. The kernel density estimation method is used to estimate the probability distribution. We have used a Gaussian kernel with bandwidth of five to smoothen the curve. The mean value of eccentricity is also calculated for each popularity class.

*Implementation details:* Python 3.8 was used to implement all the data analysis procedures of the project. To embed text ideas into numerical space, Gensim Doc2Vec library[55] was used. We have used the Distributed Bag of Words (DBOW) model in the Doc2Vec method. The Doc2Vec parameters are different between the online experiment dataset and the GAB dataset as the size of training data is different. For the GAB dataset, the document vector size is 300 and the min_count parameter is set to 10; meanwhile, for the online experiment dataset, the document vector size is 90 and the min_count parameter is set to 7. The NetworkX library[56] was used to create and maintain collaboration networks.

## Acknowledgement

This work was supported in part by the US National Science Foundation under Grant 1734147 and the JSPS KAKENHI Grant 19K21571.

## Contributions

S.P. and H.S. conceptualized the study and wrote the manuscript. S.P. carried out implementation of study, social media data collection, data analysis, data visualization and figures preparation. Y.C., Y.D., M.K., N.G.M, S.D.D., F.J.Y. and H.S. contributed to data collection from human experiments. All authors contributed to manuscript review.

## Competing interests

We declare that none of the authors have competing financial or non-financial interests.

## Material and correspondence

Correspondence and requests for materials should be addressed to Sriniwas Pandey or Hiroki Sayama.

## Human Research Participants

Experiments were conducted after an approval from the Institutional review board at Binghamton University, NY, USA.

## Data and code availability

Code can be made available upon request. Due to the IRB restrictions supporting data is not available.